\documentclass[aps,prd,floatfix,showpacs,nofootinbib,twocolumn,twoside]{revtex4}
\usepackage{amsmath} 
\usepackage{amssymb}
\usepackage{hyperref}
\usepackage{epsfig}
\usepackage{cancel}
\usepackage{epstopdf}
\usepackage{pstricks}
\usepackage{graphics}
\usepackage{graphicx}
\usepackage{multirow}

\newcommand{\GeV}{\text{GeV}}
\newcommand{\tev}{\text{TeV}}
\newcommand{\invfb}{\text{fb}^{-1}}

\newcommand{\etmiss}{\cancel{E}_T}
\newcommand{\be}{\begin{equation}}
\newcommand{\ee}{\end{equation}}
\newcommand{\bea}{\begin{eqnarray}}
\newcommand{\eea}{\end{eqnarray}}

\begin{document}

\setlength{\pdfpageheight}{\paperheight}
\setlength{\pdfpagewidth}{\paperwidth}

\title{Probing dark matter couplings to top and bottom at the LHC}

\author{Tongyan Lin, Edward W. Kolb, and Lian-Tao Wang}

\affiliation{Kavli Institute for Cosmological Physics and the Enrico
  Fermi Institute, The University of Chicago, 5640 S. Ellis Ave,
  Chicago, Il 60637}

\begin{abstract}
Monojet searches are a powerful way to place model-independent
constraints on effective operators coupling dark matter to the
standard model. For operators generated by the exchange of a scalar
mediator, however, couplings to light quarks are suppressed and the
prospect of probing such interactions through the inclusive monojet
channel at the LHC is limited. We propose dedicated searches, focusing
on bottom and top quark final states, to constrain this class of
operators.  We show that a search in mono $b$-jets can significantly
improve current limits. The mono-$b$ signal arises partly from direct
production of $b$-quarks in association with dark matter, but the
dominant component is from top quark pair production in the kinematic
regime where one top is boosted.  A search for tops plus missing
energy can strengthen the bounds even more; in this case signal and
background have very different missing energy distributions. We find
an overall improvement by several orders of magnitude in the bound on
the direct detection cross section for scalar or pseudoscalar
couplings.
\end{abstract}

\date{\today}

\maketitle

\section{Introduction}

Production and detection of dark matter is one of the most exciting
new physics opportunities at the \textit{Large Hadron Collider} (LHC).
The strategy to search for dark matter (DM) depends on the physics in
the yet-to-be fully explored energy range of the LHC. In the maverick
scenario \cite{Beltran:2010ww}, the DM is the only new particle
produced and all other new particles are beyond the scale of the
LHC. Then the interaction of the DM with standard model (SM) particles
at these energies can be described in terms of an effective field
theory (EFT).

In this case the DM signal at the LHC is missing transverse energy
($\etmiss$) signals such as monojets
\cite{Beltran:2010ww,Goodman:2010yf,Goodman:2010ku,Fox:2011pm,Fox:2012ee,Rajaraman:2011wf}
or monophotons \cite{Fox:2011fx,Fox:2011pm}. With an EFT description,
one can classify all relevant interactions at the LHC in a
straightforward way. This scenario also has the advantage that the
connection between DM annihilation, direct detection, and collider
signals is simple.

The ATLAS \cite{ATLAS:2012ky,ATLAS-CONF-2012-147} and CMS
\cite{Chatrchyan:2012me,CMS-PAS-EXO-12-048} collaborations have
published monojet constraints on the scale of new interactions in the
EFT, which are then used to place constraints on the DM-nucleon
scattering cross section. These constraints are most effective for DM
masses below 100 GeV. Meanwhile, there has been rapid progress in the
direct detection of dark matter \cite{Aprile:2012nq,Ahmed:2010wy},
with the strongest bounds on DM-nucleon scattering at DM mass of
around $50\ \GeV$ and for spin-independent scattering. These two
approaches are complementary, and connecting them has been the focus
of many recent studies \cite{Shoemaker:2011vi,Frandsen:2012rk}.

\begin{figure*}[tbh]
 \includegraphics[width=0.3\textwidth]{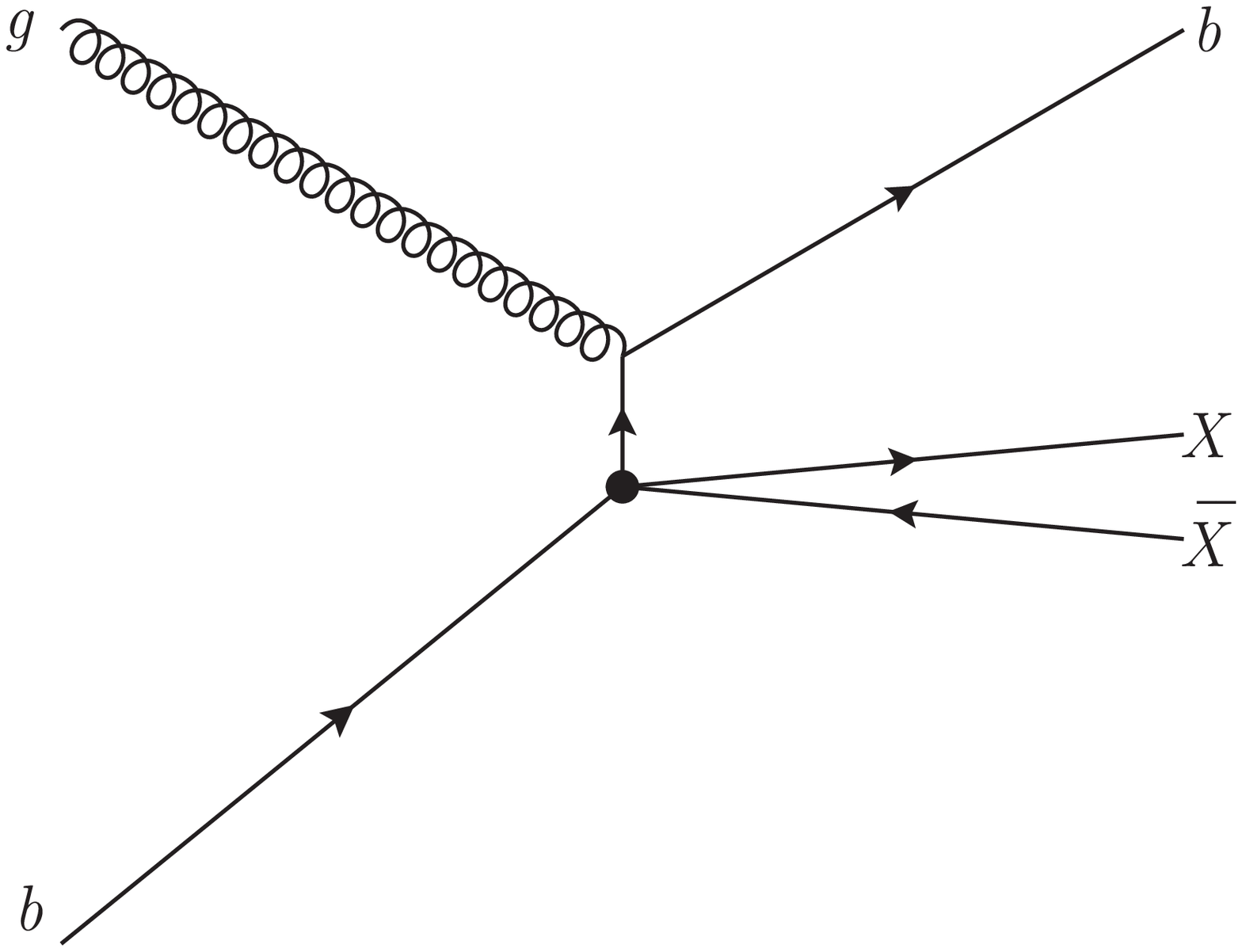}\hspace{.2cm}
 \includegraphics[width=0.3\textwidth]{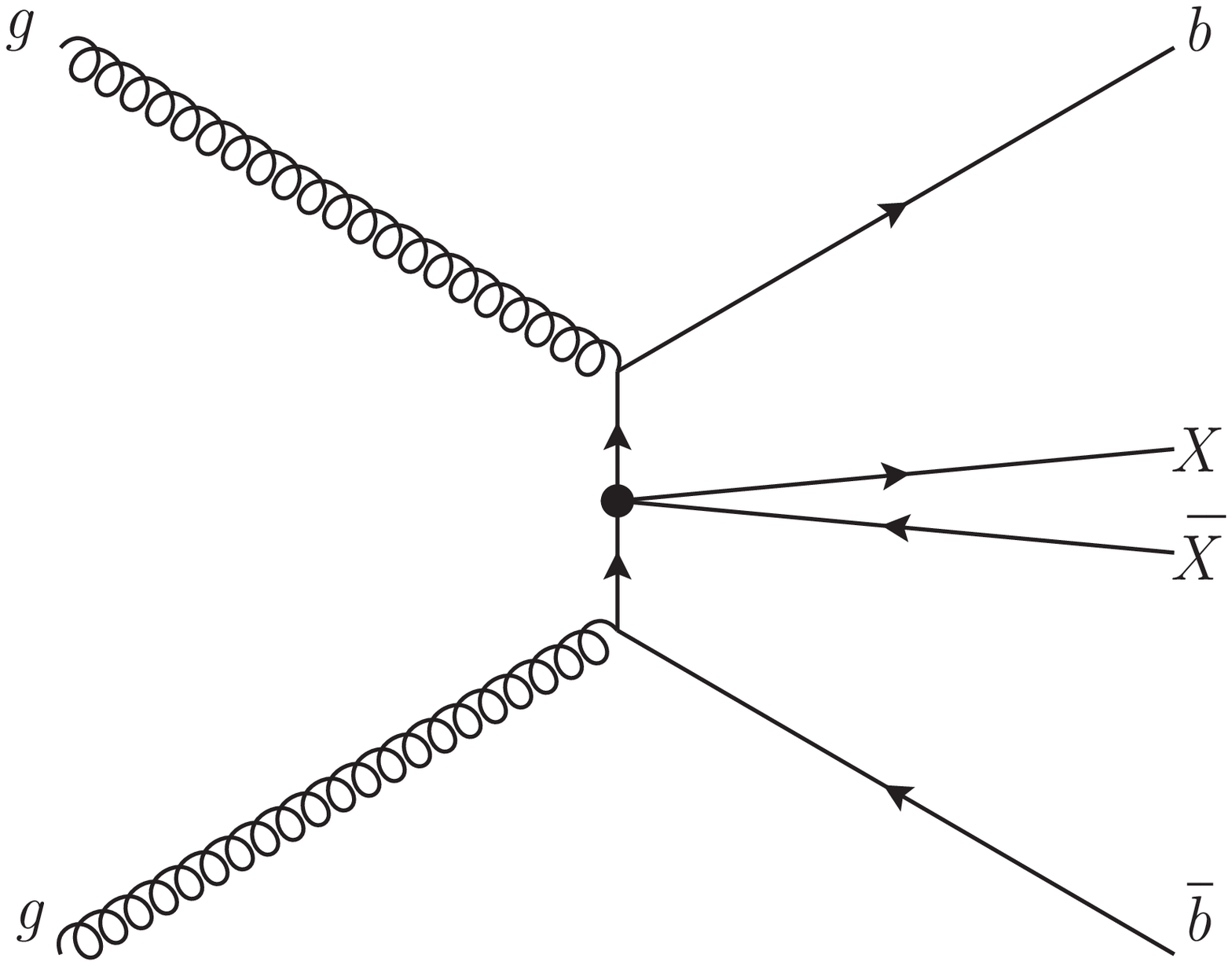}\hspace{.2cm}
 \includegraphics[width=0.3\textwidth]{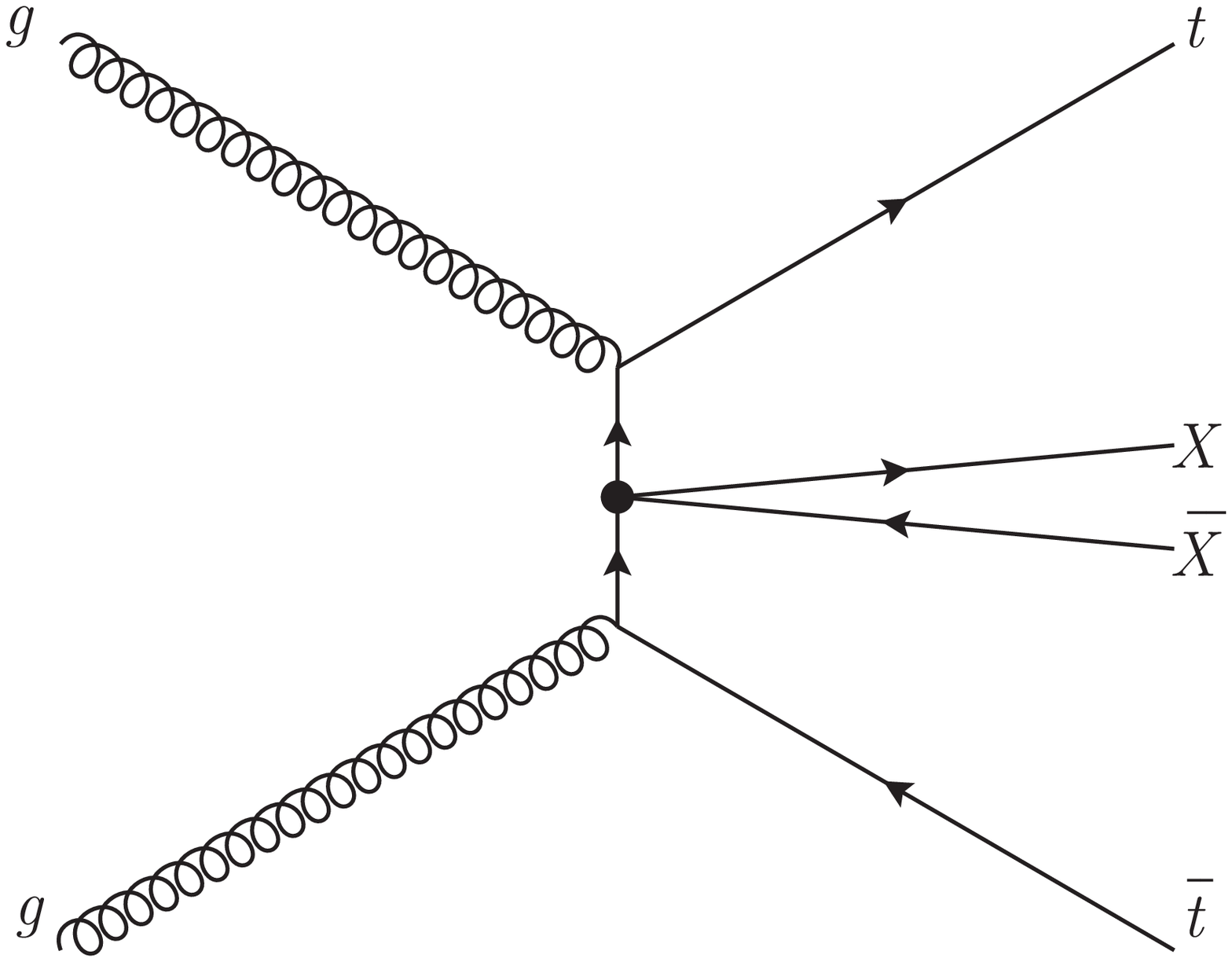}
\caption{\label{fig:diagrams} Some of the dominant diagrams
  contributing to associated production of DM with bottom and top
  quarks.}
\end{figure*}

\subsection{Scalar Operator}

While the monojet search is extremely effective for many of the
possible operators, it is not necessarily the optimal way to study all
of them. In particular, it is challenging to constrain the scalar
operator, where interactions between dark matter and quarks are
mediated by a heavy scalar mediator: \be
\label{eq:op_scalar}
      {\cal O} = \frac{ m_q}{M_*^3}\bar q q \bar X X,
\ee
summing over all quarks.\footnote{There are closely related operators,
  for example the pseudoscalar operator $\frac{ m_q}{M_*^3}\bar q
  \gamma^5 q \bar X \gamma^5 X$, which will have almost identical
  collider constraints.  The direct detection cross section for these
  operators is velocity suppressed, however, so the best limits will
  come from the LHC. It should be understood that our limits apply to
  both scalar and pseudoscalar operators.}  The form of the
interaction is fixed by minimal flavor violation (MFV)
\cite{D'Ambrosio:2002ex}. Scalar interactions with SM quarks are
typically strongly constrained by flavor changing neutral current
measurements, but in MFV these dangerous flavor violating effects are
automatically suppressed.

Because the interactions are proportional to quark mass, however, the
monojet$+ \etmiss$ signal rate appears to be suppressed by the light
quark masses.  The ATLAS monojet search based on $4.7\ \invfb$ at 7
\tev\ sets a limit of $M_* > 30\ \GeV$ \cite{ATLAS:2012ky}, including
only couplings to charm and lighter quarks. This bound is much weaker
than constraints on operators mediated by vector or axial
interactions.

In this paper, we point out that the direct search for production of
dark matter in association with third generation quarks can enhance
the reach of the LHC for dark matter coupled to quarks through a
scalar interaction.  Direct $b$ production gives rise to a mono
$b$-jet signal. We also show that the kinematics of top quark pairs
plus dark matter is such that boosted tops may form the dominant
contribution to the mono-$b$ signal. However, monojet searches veto on
more than 2 hard jets, so a better strategy to probe the couplings to
top quarks is the study of $t \bar t + \etmiss$ final states.

The scalar interaction has also been studied recently in
Ref.~\cite{Haisch:2012kf}, which showed that heavy quarks in loops can
significantly enhance inclusive monojet production. We focus instead
on direct identification of the heavy quarks in the final state.
Mono-$b$ final states from dark matter have also been studied in Refs.
\cite{Bhattacherjee:2012ch,Cheung:2012xb}, although not in the context
of MFV, so the top quark contribution to the mono-$b$ signal was
not considered.

In Section~\ref{sec:monob} we study the mono $b$-jet signal where the
leading jet is $b$-tagged. This search can improve constraints on the
DM-nucleon cross section, $\sigma_n$, by several orders of magnitude
compared to current ATLAS limits.  In Section~\ref{sec:ttmet} we show
an even stronger limit can be obtained from a search for dark matter
in association with top quarks, $t \bar t + \etmiss$.  This is also
the final state studied in searches for stops, supersymmetric partners
to tops, and we use published results to derive limits. We find that
the limit on $\sigma_n$ is stronger by another factor of approximately
2 compared to the mono $b$-jet search.

\section{Mono $b$-jet search \label{sec:monob}}

The scalar operator gives rise to $b$-jets plus $\etmiss$ via direct
$b$ production, as well as from production of top quarks which then
decay. Direct $b$ production occurs through $b$ and gluon-initiated
processes, such as $bg \to \bar X X + b$; several example diagrams are
shown in Fig.~\ref{fig:diagrams}.
  In comparison to the light quark initial states, 
these processes are suppressed by the $b$-quark parton
density. However, the enhancement due to the MFV form for the coupling
is more than enough to compensate this.

Furthermore, $gg \to \bar X X + t\bar{t}$ turns out to be the dominant
contribution to the monojet signal. Thus, the final states are highly
$b$-enriched.  At the same time, focusing on exclusive $b$-tagged
final state reduces the SM backgrounds significantly. Therefore, we
expect an improvement in the LHC reach for the scalar operator by
requiring a $b$-tagged monojet.

Before presenting our results, we summarize our event simulation
methods. We use MadGraph 5 \cite{Alwall:2011uj} for parton-level cross
sections, interfaced to Pythia 6 \cite{Sjostrand:2006za} for showering
and hadronization, and Delphes 2 \cite{Ovyn:2009tx} for detector
simulation. For Delphes, we set a 60$\%$ tagging efficiency for $b$,
10$\%$ mistag for $c$, and $0.2 \%$ mistag rate for light quarks and
gluons \cite{Aad:2009wy}. Jets are clustered into $R=0.4$ anti-$k_T$
jets.

Up to two hard jets are allowed in the monojet and mono-$b$ searches,
so we must consider Next-to-Leading-Order (NLO) corrections in our
simulation of the signal. We generate matched samples with $k_T$-jet
MLM matching. 
For SM backgrounds, we generate $W/Z$ and
$t{\bar t}$ with up to 2 jets. For the signal, we generate $X{\bar X}
+ $jets, including up to 2 jets, for all flavors other than tops. We
separately include $X{\bar X}+t{\bar t}$ at leading order. Finally, we
normalize all matched samples with NLO cross sections computed using
MCFM \cite{Fox:2012ru}.

For the signal region we require $\etmiss > 350\ \GeV$, a leading
$b$-tagged jet with $p_T > 100\ \GeV, |\eta| < 2.5$, and no isolated
leptons. We also allow an additional softer jet, but no more than two
jets with $p_T > 50\ \GeV$. There is a cut on the azimuthal separation
between $\etmiss$ and the second jet, $\Delta \phi(\etmiss,p_T^{j_2}) >
0.4$, in order to suppress the mismeasured dijet background. This signal
region overlaps well with those used in previous studies, and
furthermore the dependence on the cut values appears to be mild.

The resulting cross sections at 8 TeV are given in
Table~\ref{tab:monob_8TeV}. We have split the signal into three
contributions: coupling to charm and light quarks, direct $b$ and
$b{\bar b}$ production from coupling to $b$, and $t {\bar t}$
production. 

Associated production of DM with $t{\bar t}$ constitutes the dominant
signal for both monojet and mono-$b$ signals because of the
enhancement from the top mass and because of the production of boosted
tops which can be tagged as $b$-jets.  Events where only one top is
boosted, and where the other top gives rise to low $p_T$ jets, can
pass the mono-$b$ cuts.  Of the events that pass the $\etmiss$
requirement and lepton veto, $17\%$ of events survive the veto based
on the $p_T$ of the third jet. In comparison, 80$\%$ of events from
direct $b$ production survive the jet veto.  In both $b$ and top
production, about 50$\%$ of those events then have a leading jet which
is $b$-tagged. Note that this assumes the same $b$-tagging
efficiencies for the $b$-jets inside the boosted tops. Without dedicated
study by the experimental collaborations, this is an idealized
assumption. We note that most of the top jets are from mildly boosted
tops, with have $p_T \approx 400\ \GeV$. Hence, we do not expect the
$b$-tagging efficiency to degrade significantly.  At the same time, it
may be possible to use additional information to tag these jets as
coming from boosted tops. We will leave this for a future study.

The dominant SM backgrounds are $Z(\nu \bar \nu) + $jets and $W+$jets,
where a jet fakes a $b$-jet. Although these are suppressed by the mistag
rate, they are still a larger contribution than direct $b$ production
from $Z+b$ and $t{\bar t}$ backgrounds. Depending on the $b$-tagging
algorithms used, however, it may be possible to further reduce the
background from $Z/W$+jets.

\begin{figure}[tb]
 \includegraphics[width=0.47\textwidth]{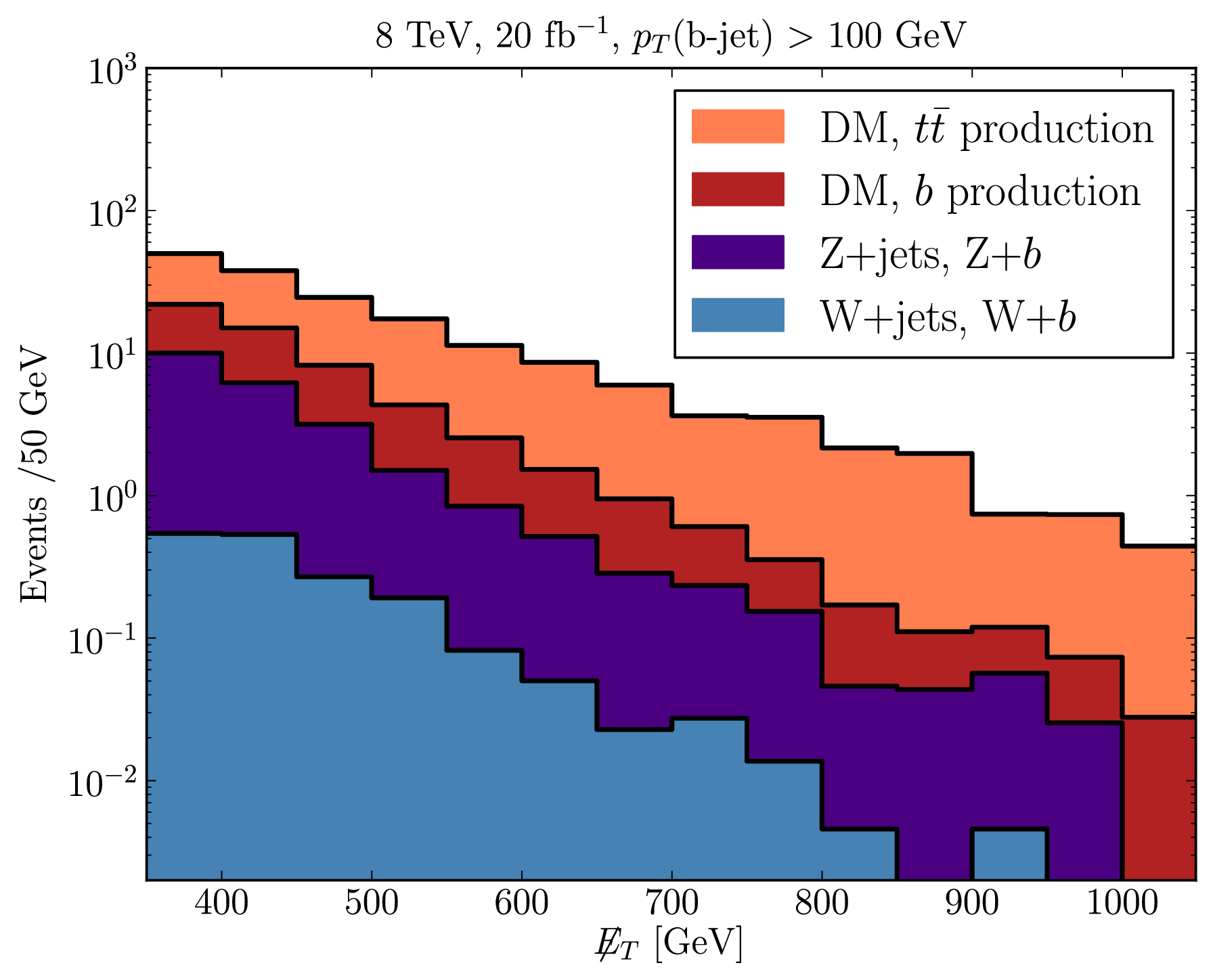}\\
 \includegraphics[width=0.47\textwidth]{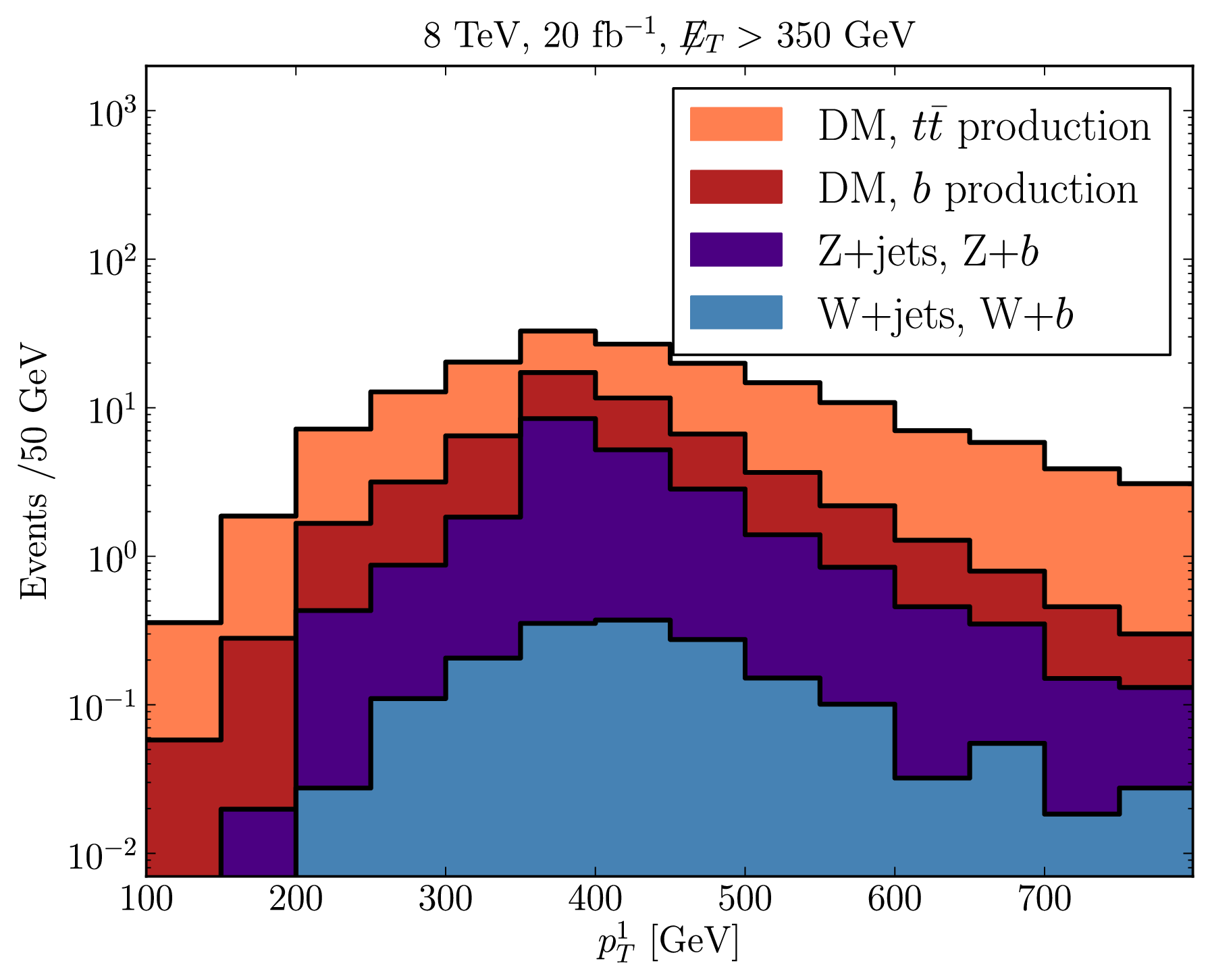}
\caption{\label{fig:monobjet}{\bf Mono-$b$ search at 8 TeV: }
  Distributions for $\etmiss$ and $p_T^{j_1}$, the transverse momentum
  of the leading $b$-jet, for some of the dominant SM backgrounds and
  the signal. We separate the DM signal into contributions from direct
  $b$ production and from $t\bar{t}$ production. For the signal we
  take $M_* = 50\ \GeV$ and $m_X= 10\ \GeV$. }
\end{figure}

Kinematic distributions in $\etmiss$ and leading jet $p_T$ are shown
in Fig.~\ref{fig:monobjet}. The $\etmiss$ spectrum for $b$ production
is very similar to that for $Z$+jets, despite the fact that the signal
arises from a contact interaction. This is partly because in the
signal case, the initial states include sea quarks, while for
the dominant $Z$+jets background roughly 70$\%$ of events are
initiated by at least one valence quark. Meanwhile, $X\bar{X} +
t\bar{t}$ final states tend to have more $\etmiss$ because of the
requirement of producing massive top quarks.

\begin{table}[tb]
\begin{center}
\begin{tabular}{|c|c|c|c|c|}
\hline
& Process & Monojet & $b$-tag  & $b$-tag on $j_1$   \\
\hline
\multirow{4}{*}{Background} &
$Z$+jets(fake) & 406 fb & 11 fb & 7 fb \\
& $Z$+$b$+jet      & 6.7 fb & 4 fb  & 3 fb \\
& $W$+jets,$W$+$b$      & 95 fb  & 3 fb  & 2 fb  \\
& $t\bar{t}$+jets     & 16 fb & 11 fb & 6 fb \\
\hline
\multirow{3}{*}{Signal} &
$\bar X X + $jets & 11 fb & 0.9 fb &  0.7 fb\\ 
& $\bar X X + b + $jets & 65 fb & 40 fb &  33 fb\\ 
& $\bar X X + t{\bar t}$ & 244 fb & 156 fb &  113 fb\\ 
\hline
\end{tabular}
\end{center} 
\caption{{\bf Monojet and mono-$b$ search at 8 TeV: } Cross sections
  for dominant backgrounds and signal with cuts of $\etmiss >
  350\ \GeV$, $p_T^{j_1} > 100\ \GeV$ as described in the text.  For
  the signal we take $M_* = 50\ \GeV$ and $m_X= 10\ \GeV$. The row
  {$\bar X X +$ jets} includes only DM coupling to charm and lighter
  quarks. Note that {$\bar X X + b + $ jets} includes single $b$-jet
  and $b{\bar b}$ production. In the column labeled $b$-tag, a $b$-tag
  on any jet with $p_T > 50\ \GeV$ is required, while in the last
  column the leading jet must be $b$-tagged; this choice does not lead
  to significantly different results in setting limits.}
\label{tab:monob_8TeV}
\end{table}

To estimate the expected bound, we compute the number of signal events
such that $\chi^2 < 2.71$ to obtain a 90\% CL bound
\cite{Fox:2011pm}. A systematic uncertainty of 5$\%$ is assumed. We
also compute bounds for 14 TeV and 100~$\invfb$, keeping the same
cuts. A higher $\etmiss$ cut can improve bounds in the inclusive
monojet case \cite{Rajaraman:2011wf}; in this case, however, stronger
cuts would also require higher $p_T$ $b$-jets, where the $b$-tagging
efficiency can degrade.

\begin{figure*}[th]
 \includegraphics[width=0.47\textwidth]{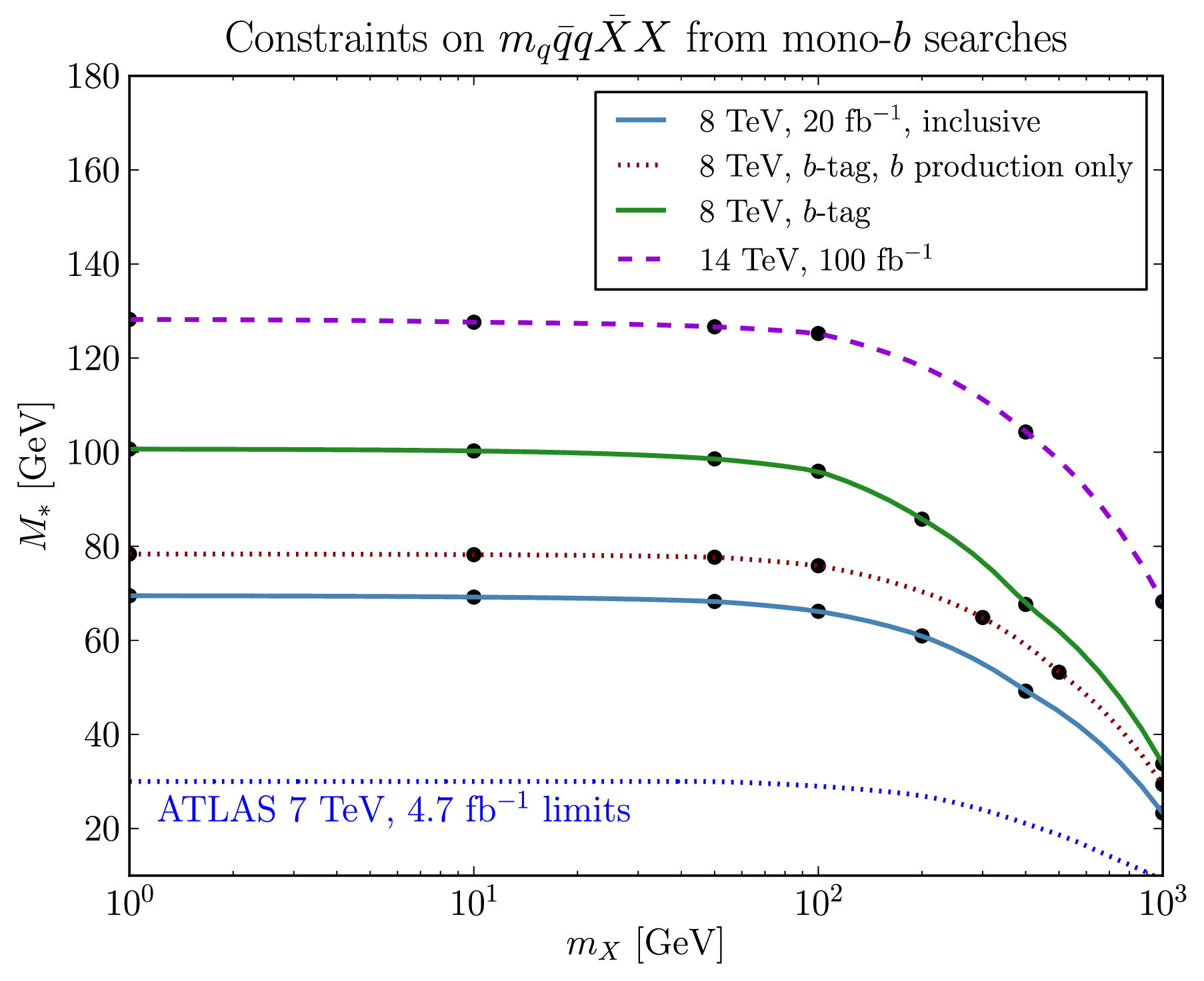}
 \includegraphics[width=0.47\textwidth]{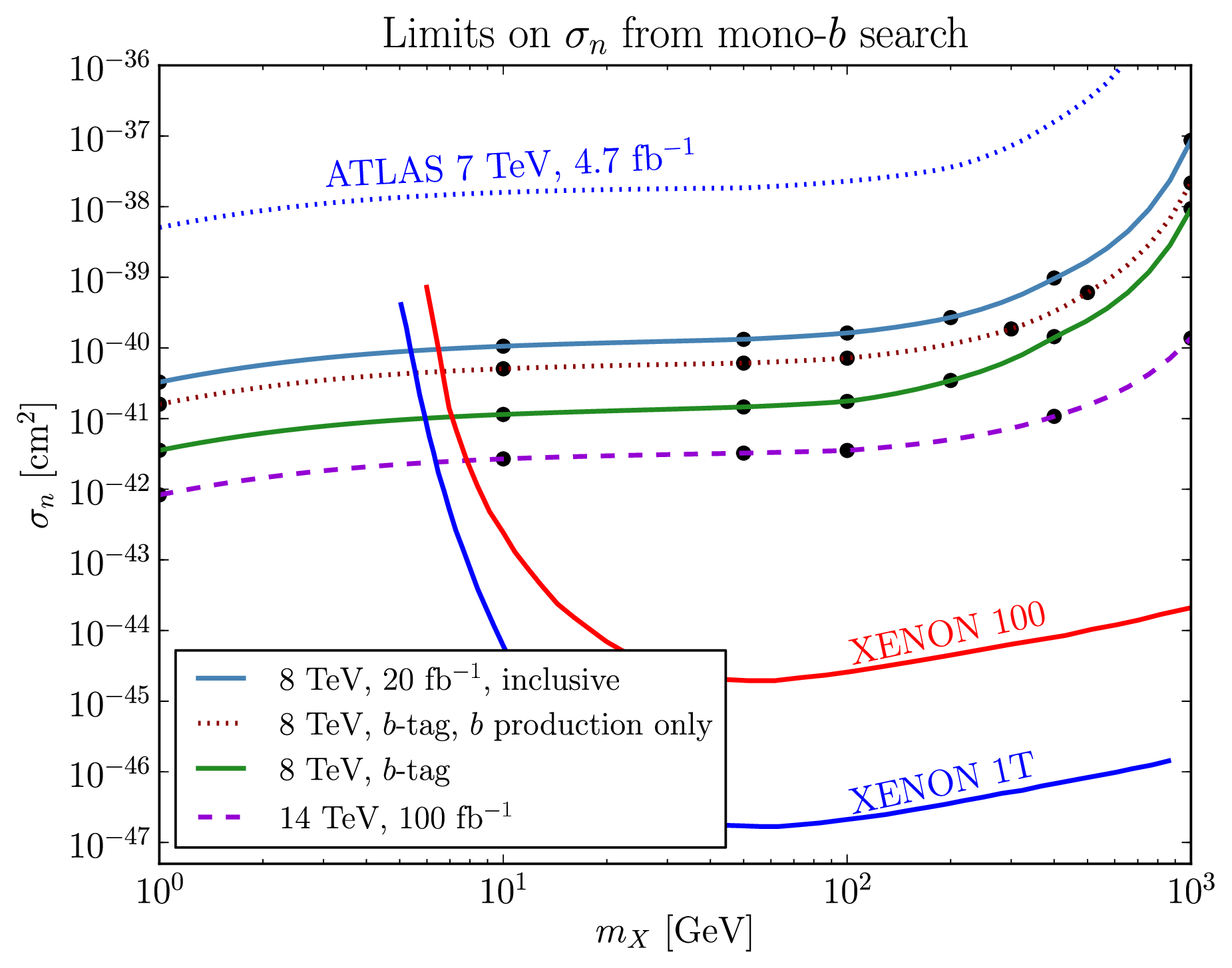}
\caption{\label{fig:monob_limits} {\it (Left)} Expected 90$\%$ CL
  limits on the scalar operator from a mono-$b$ search, including
  couplings to tops and bottoms. For the mono-$b$ search at 8 TeV we
  also show the limit if $b$-jets from top production are not included
  (dotted line). For comparison we include limits for an inclusive
  monojet search with no $b$-tag, and current ATLAS limits from
  \cite{ATLAS:2012ky}. {\it (Right)} Corresponding constraints on the
  spin-independent nucleon scattering cross section, along with
  XENON100 limits \cite{Aprile:2012nq}, and projected sensitivity for
  XENON1T \cite{Aprile:2012zx}. }
\end{figure*}

Fig.~\ref{fig:monob_limits} shows our constraints on $M_*$ in the left
panel and the corresponding limits for direct detection in the right
panel. The scalar operator gives rise to spin-independent DM-nucleon
scattering, with a cross section of
\begin{align}
  \sigma_n  = \frac{(0.38 m_n)^2 \mu_X^2}{\pi M_*^6}  \approx 2
  \times 10^{-38} \text{cm}^2 \left( \frac{30\ \GeV}{M_*} \right)^6,
\end{align}
compared to XENON100 limits of about $10^{-43} \text{cm}^2$ at $m_X =
10\ \GeV$ \cite{Aprile:2012nq}.  We find a factor of 8 improvement in
$\sigma_n$ limits with a mono-$b$ search compared to an inclusive
monojet search. Overall, we find a factor of 250 improvement compared
to an inclusive search where only the coupling to charm quarks and
lighter is considered (as in the most recent ATLAS study
\cite{ATLAS:2012ky}).


\section{Tops plus missing energy search \label{sec:ttmet}}

As shown in the previous section, the process $gg \to X \bar X+ t \bar
t$ contributes the dominant component of the monojet and the mono-$b$
signals. The monojet and mono-$b$ searches veto on more than two
high-$p_T$ jets, however, cutting out a large fraction of $t\bar t$
events. A stronger constraint on this coupling can be obtained from
dedicated searches.

Models of supersymmetry also have a signature of top pairs plus
missing transverse energy. We apply the recent ATLAS 8 TeV search for
top-quark superpartners with 1-lepton final states
\cite{ATLAS-CONF-2012-166} using 13 fb$^{-1}$ of data to these scalar
dark-matter couplings.\footnote{We have also calculated constraints
  using the CMS 1-lepton final state search \cite{CMS-PAS-SUS-12-023}
  and obtain limits that are similar although slightly weaker.} The
signal regions require 1 isolated lepton, $\etmiss >$ 150 GeV,
transverse mass\footnote{The transverse mass is defined as $(m_T)^2 =
  2 p_T^{lep} \etmiss ( 1- \cos \Delta \phi )$ with $\Delta \phi$ the
  azimuthal separation between lepton and missing momentum
  directions.} $m_T > 120\ \GeV$, 4 jets with $p_T >
(80,60,40,25)\ \GeV$ and at least 1 $b$-tag.

\begin{figure*}[tb]
 \includegraphics[width=0.47\textwidth]{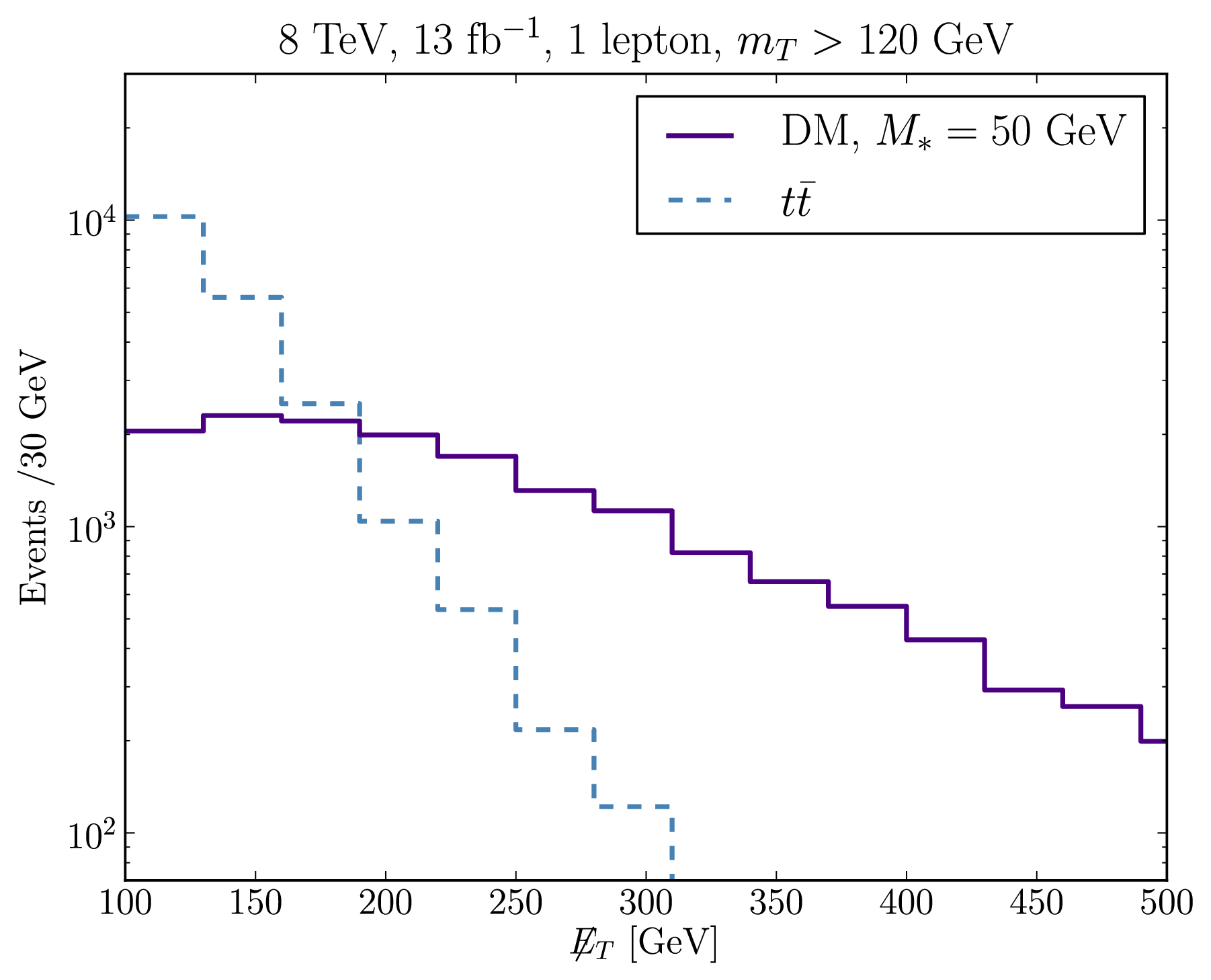}
 \includegraphics[width=0.47\textwidth]{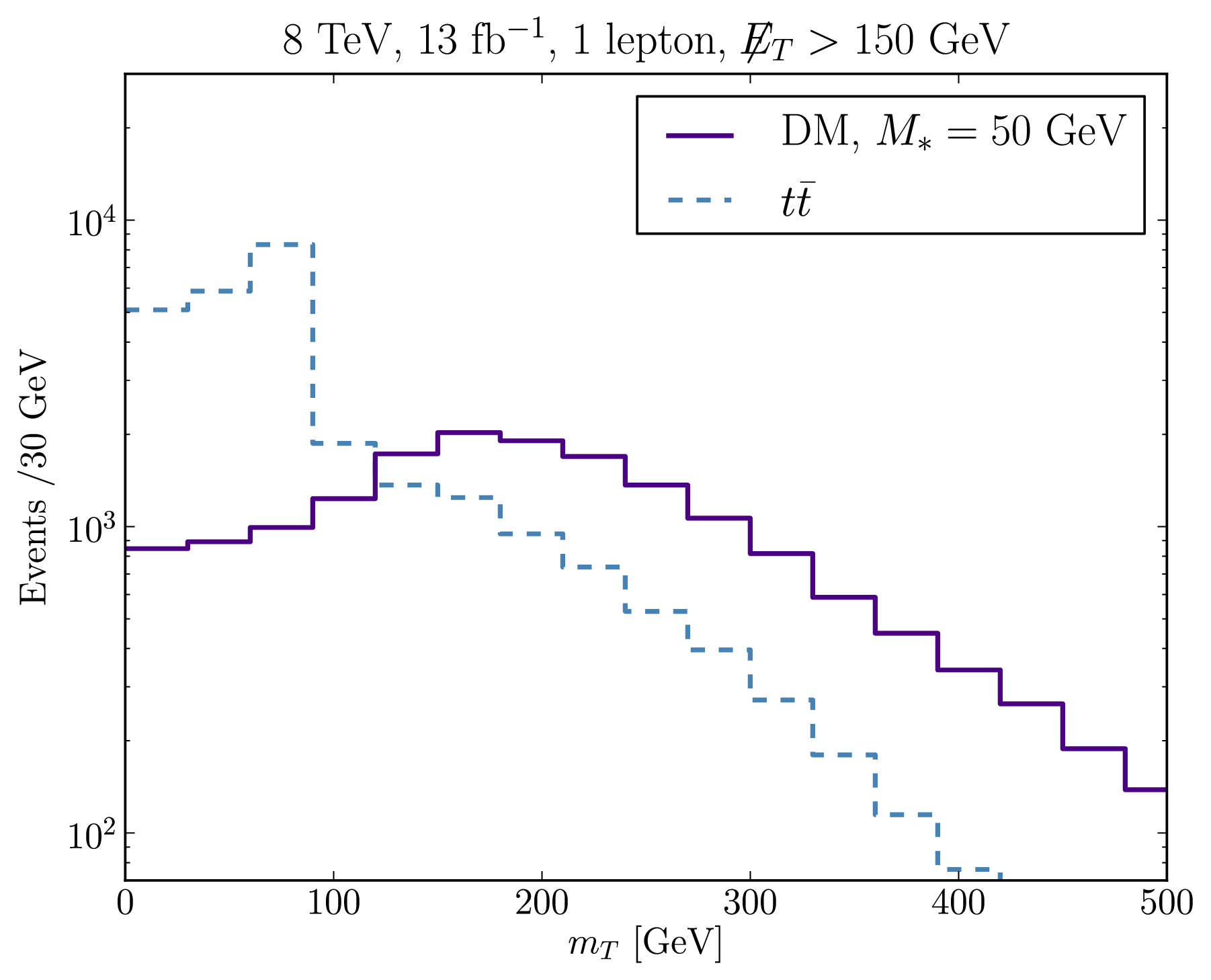}
\caption{\label{fig:stop_1L} ({\it Left}) $\etmiss$ distribution after
  requiring an isolated lepton and $m_T > 120\ \GeV$. ({\it Right})
  Transverse mass $m_T$ distribution requiring an isolated lepton and
  $\etmiss > 150\ \GeV$. The dark matter mass is $m_X = 10\ \GeV$.}
\end{figure*}

Fig.~\ref{fig:stop_1L} shows the $\etmiss$ and $m_T$ distributions of
the signal and the dominant background, $t\bar{t}$. The DM signal is
significantly harder in the $\etmiss$ spectrum, whereas the background is
highly peaked towards low $\etmiss$ because the primary source of
$\etmiss$ is from the neutrinos in the top decay. Meanwhile, it is
unlikely that stronger cuts on $m_T$ above 120 GeV would substantially
improve the ratio of signal to background.

\begin{figure*}[tbh]
 \includegraphics[width=0.47\textwidth]{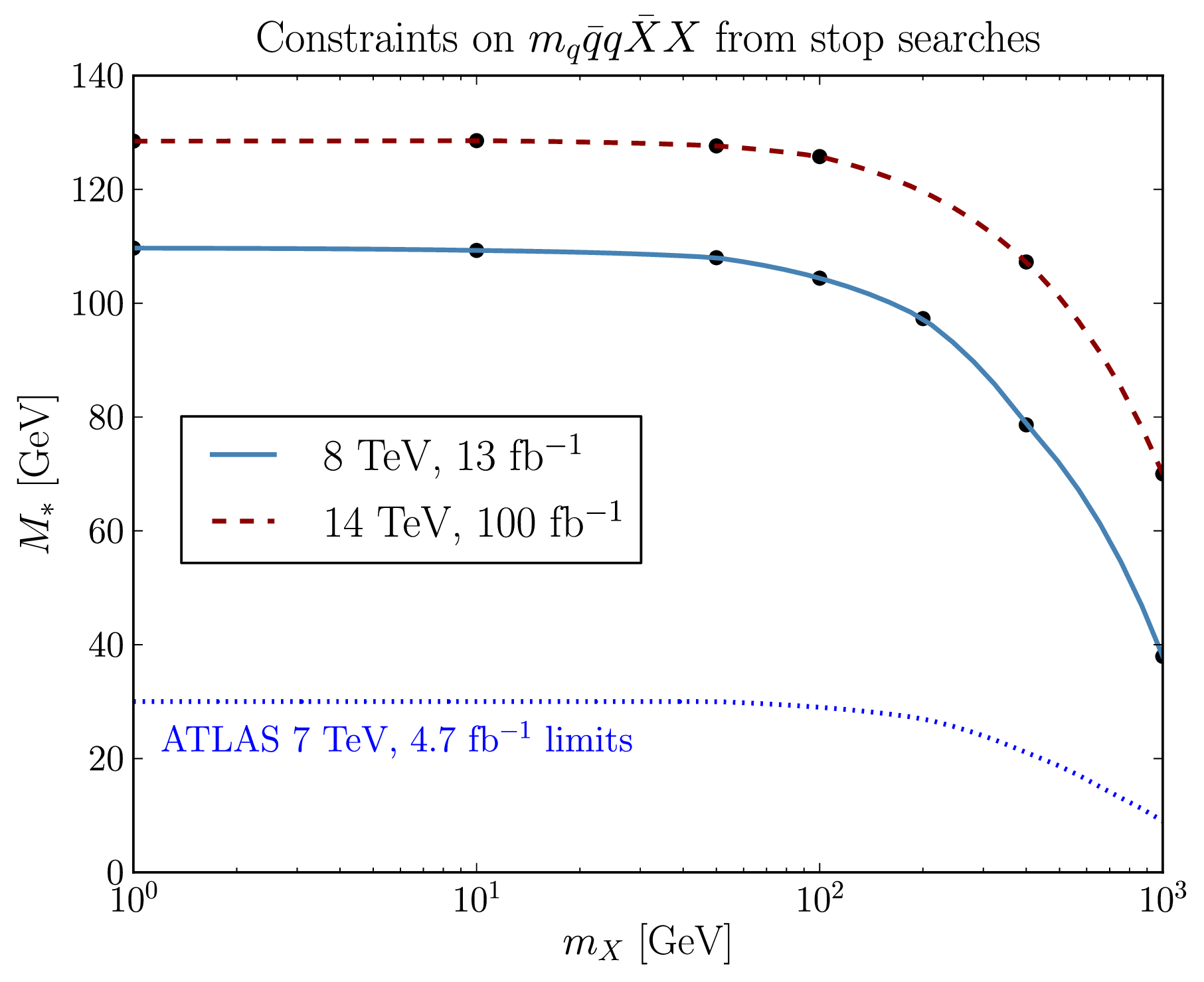}
 \includegraphics[width=0.47\textwidth]{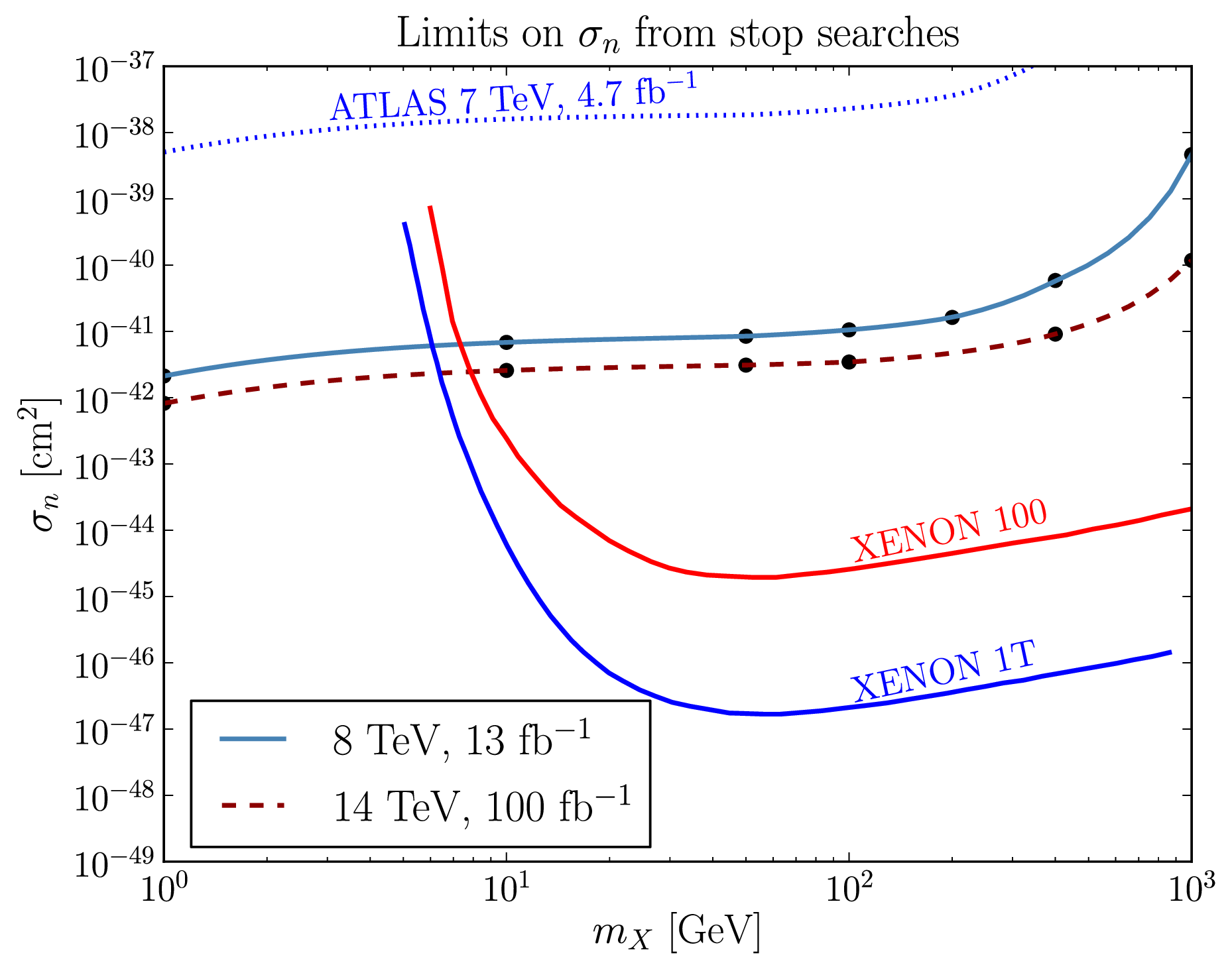}
\caption{\label{fig:stop_limits} {\it (Left)} Expected 90$\%$ CL
  limits on scalar operator from applying a search for supersymmetric
  tops with one lepton in the final state. 8 TeV limits are obtained
  using the results of \cite{ATLAS-CONF-2012-166}. Also shown are
  ATLAS limits from \cite{ATLAS:2012ky}. {\it (Right)} Corresponding
  constraints on nucleon scattering cross section, along with XENON100
  limits \cite{Aprile:2012nq}, and projected sensitivity for XENON1T
  \cite{Aprile:2012zx}. }
\end{figure*}

We find the best constraints come from the signal region D (SRD) of
the ATLAS study, with $\etmiss >$ 225 GeV. Although there is another
signal region with $\etmiss >$ 275 GeV, the systematic uncertainties increase
significantly. We thus apply the ATLAS SRD cuts to simulated data to
derive our 13 $\invfb$ limits.  The signal cross section with these
cuts is
\begin{equation}
  \sigma_{\rm{signal}} =   173\ {\rm fb}
\end{equation}
assuming $M_* = 50\ \GeV$ and $m_X=10\ \GeV$. We give limits on $M_*$
and $\sigma_n$ in Fig.~\ref{fig:stop_limits}. Because uncertainties
are systematics dominated for this signal region, we do not expect a
significant improvement of limits with 20 $\invfb$ of data.

Fig.~\ref{fig:stop_limits} also shows limits for 14 TeV with 100
$\invfb$ of data, keeping the same cuts as above. We simulate
$W$+jets in addition to $t\bar{t}$ for our background estimate, and assume the systematic error
on the background is the same as in the 8 TeV analysis. We also
calculated constraints for a search with an all-hadronic final state
\cite{ATLAS-CONF-2013-024,CMS-PAS-SUS-12-028}; in this case it may be
possible to improve the bounds on $M_*$ by 10-20$\%$, depending on the
detector acceptances and systematic uncertainties.


\section{Discussion}

We have shown that limits on scalar (and pseudoscalar) interactions of
dark matter with quarks can be improved significantly by directly
searching for final states with $b$-jets and tops. Compared to an
analysis including only light quarks, we find a factor of 400
improvement in limits on $\sigma_n$, and compared to an inclusive
monojet search including couplings to all quarks, we find a factor of
15 improvement. For 8 TeV data, the corresponding constraints on
direct detection are below the regions favored for light dark matter
interpretations of DAMA \cite{Bernabei:2010mq} and CoGeNT
\cite{Aalseth:2011wp}.


Couplings to heavy quarks can also lead to an enhancement of inclusive
monojet production through loops; Ref.~\cite{Haisch:2012kf} found $M_*
> 148^{+12}_{-11}\ \GeV$ for small DM mass using 7 TeV data. However,
these loop corrections assume that the operator is generated by a
heavy neutral scalar.  Although our constraints are weaker, the
searches discussed here directly probe couplings of dark matter to top
and bottom. Furthermore, the $\etmiss$ spectrum for the $t\bar{t}$
final state is strikingly different from the background. It may be
possible to use the difference in shapes to improve limits from the
searches discussed here.

Finally, in this paper we have assumed a contact interaction for
simplicity. As discussed in Refs. \cite{Shoemaker:2011vi,Fox:2012ee},
this assumption must be compared to the derived bounds on $M_*$. In
this case, the best limit we obtain at 8 TeV is $M_* > 110\ \GeV$, and
for this value a significant fraction of events (over 50$\%$) violate
the criteria in Refs. \cite{Shoemaker:2011vi,Fox:2012ee}. A UV
completion for this operator is necessary to derive fully consistent
constraints. At the same time, the results will be more
model-dependent and we reserve this analysis for future work.

\begin{acknowledgments}
We thank Patrick Fox, David Krohn, Bjoern Penning, Brian Shuve, and Ciaran
Williams for helpful discussions, and in particular Bjoern Penning for
his constant feedback. Paddy and Ciaran provided an advance copy and
assistance with MCFM. E.W.K. thanks the Physics Department of the
University of Rome ``La Sapienza'' and INFN, Sezione di Padova, where
part of this work was completed. L.T.W. is supported by the NSF under
grant PHY-0756966 and the DOE Early Career Award under grant
de-sc0003930. T.L. is grateful to Perimeter Institute for their
hospitality as this paper was being finished. Research at the
Perimeter Institute is supported in part by the Government of Canada
through Industry Canada, and by the Province of Ontario through the
Ministry of Research and Information (MRI).
\end{acknowledgments}

\bibliography{monojet}

\end{document}